# Sparse Principal Component based High-Dimensional Mediation Analysis

YI ZHAO*, MARTIN A. LINDQUIST, BRIAN S. CAFFO

*Department of Biostatistics, Johns Hopkins Bloomberg School of Public Health*

*615 N Wolfe St, Baltimore, MD, USA*

yzhao99@jhmi.edu

Summary

Causal mediation analysis aims to quantify the intermediate effect of a mediator on the causal pathway from treatment to outcome. With multiple mediators, which are potentially causally dependent, the possible decomposition of pathway effects grows exponentially with the number of mediators. Huang and Pan (2016) introduced a principal component analysis (PCA) based approach to address this challenge, in which the transformed mediators are conditionally independent given the orthogonality of the PCs. However, the transformed mediator PCs, which are linear combinations of original mediators, are difficult to interpret. In this study, we propose a sparse high-dimensional mediation analysis approach by adopting the sparse PCA method introduced by Zou *and others* (2006) to the mediation setting. We apply the approach to a task-based functional magnetic resonance imaging study, and show that our proposed method is able to detect biologically meaningful results related to the identified mediator.

*To whom correspondence should be addressed.

1. Introduction

Causal mediation analysis has been widely applied in social, psychological, and biological studies to evaluate the intermediate effect of a variable (called mediator) on the causal pathway from an exposure/treatment to a target outcome and delineate the underlying causal mechanism. The single mediator setting has been extensively studied (Baron and Kenny, 1986; Holland, 1988; Robins and Greenland, 1992; Pearl, 2001; Ten Have *and others*, 2007; MacKinnon, 2008; Sobel, 2008; VanderWeele and Vansteelandt, 2009; Imai *and others*, 2010; VanderWeele, 2015). During the past decade, methods for dealing with multiple mediators has attracted increasing attention (MacKinnon, 2000; Preacher and Hayes, 2008; Imai and Yamamoto, 2013; Wang *and others*, 2013; VanderWeele *and others*, 2014; VanderWeele and Vansteelandt, 2014; Zhao *and others*, 2014; Boca *and others*, 2014; Daniel *and others*, 2015; Taguri *and others*, 2015; Nguyen *and others*, 2016; Lin and VanderWeele, 2017; Vansteelandt and Daniel, 2017; Steen *and others*, 2017; Calcagnì *and others*, 2017; Park and Kürüm, 2018). However, most of these methods are designed for dealing with relatively low-dimensional data. With the emergence of modern technologies (for example, high-throughput technologies in omics studies and neuroimaging technologies), datasets with a large number of variables are increasingly being collected. However, methodologies conducting mediation analysis with high-dimensional mediators are limited. Motivated by a genetics study, Huang and Pan (2016) proposed a principal component analysis (PCA) based approach to reduce the high-dimensional gene expression mediators to lower-dimensional independent single-mediator problems. Incorporating a regularized regression for the outcome model, Zhang *and others* (2016) introduced an independent screening approach for high-dimensional mediation analysis.

In the field of neuroimaging, studies on the impact of brain mediators on cognitive behavior are becoming increasingly popular (Caffo *and others*, 2007; Wager *and others*, 2008, 2009; Atlas *and others*, 2010; Lindquist, 2012; Atlas *and others*, 2014; Woo *and others*, 2015). Caffo *and* *others* (2007) presented an early attempt at addressing neurological images as mediators, though the analysis was conducted on univariate summaries extracted from the multivariate images. Recently, Chén *and others* (2017) proposed a mediation analysis approach that transforms the high-dimensional mediator candidates into independent directions of mediation (DMs). Under the linear structural equation modeling (LSEM) framework, these directions are ranked based on the proportion of the likelihood that account for. Zhao and Luo (2016) recently proposed a general mediation model under the LSEM framework to account for the causal dependencies between the mediators and introduced a new lasso-type penalty to directly regularize the mediation pathway effects to achieve simultaneous mediator selection and mediation effect estimation.

In both the PCA and the directions of mediation analysis, the acquired PCs and DMs are linear combinations of the original mediating variables. With nonzero loadings thieir interpretation is not always straightforward. An informal way of reducing the number of variables is to set a hard threshold, and force loadings with absolute value below the threshold to be zero. However, this can be potentially misleading, as it is not only the loadings but also the variance of the variable that governs its importance (Cadima and Jolliffe, 1995). Jolliffe *and others* (2003) introduced a modified PCA approach based on the Lasso (Tibshirani, 1996). Built on the fact that sparsifying the PC loadings can be expressed as a regression-type optimization problem, Zou *and others* (2006) introduced the sparse principal component analysis (SPCA) approach. The same idea was later implemented in canonical correlation analysis (CCA) (Witten *and others*, 2009; Witten and Tibshirani, 2009). In this study, we propose a sparse principal component of mediation (SPCM) approach to perform high-dimensional mediation analysis. This approach has two stages: (i) the first performs a PCA for high-dimensional mediation using the method proposed in Huang and Pan (2016); and (ii) the second sparsifies the loading vector using a (structured) regularization. In neuroimaging studies, spatial smoothing regularization is generally imposed to enforce spatial smoothness and yield meaningful biological interpretations (Grosenick *and others*, 2013; Liu *and* *others*, 2018). In this study, the fused lasso (Tibshirani *and others*, 2005) as a special case of the generalized lasso (Tibshirani and Taylor, 2011) will be employed to impose local smoothness and constancy.

This paper is organized as follows. Section 2 introduces the PCA based mediation approach for multiple mediators proposed in Huang and Pan (2016). In Section 3, we present the sparse principal component of mediation approach and apply to a task-based functional magnetic resonance imaging (fMRI) study in Section 5. Section 6 summarizes this paper with discussions.

## 2. Causal Mediation Analysis with Multiple Mediators

Mediation analysis aims to quantify the causal effect of a treatment/exposure ($X$) on the outcome ($Y$) mediated by a third variable, called the mediator ($M$). This causal relationship can be represented using a causal diagram as in Figure 1a. Linear structural equation modeling (LSEM) is a popular approach to perform mediation analysis. Let $M(x)$ and $Y(x, M(x))$ denote the potential outcome of the mediator and the outcome under treatment assignment $x$ (Rubin, 1978, 2005), the mediation models are written as

$$M(x) = \alpha_0 + \alpha x + \epsilon, \tag{2.1}$$

$$Y(x, M(x)) = \beta_0 + \gamma x + \beta M(x) + \eta, \tag{2.2}$$

where $\epsilon$ and $\eta$ are model errors with mean zero. The average total treatment effect is decomposed as

$$\begin{aligned}
\text{ATE}(x, x^*) &= \mathbb{E}[Y(x, M(x))] - \mathbb{E}[Y(x^*, M(x^*))] \\
&= \mathbb{E}[Y(x, M(x)) - Y(x, M(x^*))] + \mathbb{E}[Y(x, M(x^*)) - Y(x^*, M(x^*))] \\
&= \text{AIE}(x, x^*) + \text{ADE}(x, x^*),
\end{aligned} \tag{2.3}$$

where $x$ and $x^*$ are two distinct treatment assignments. Under models (2.1) and (2.2), $\text{ADE}(x, x^*) = \gamma(x - x^*)$ is the average (controlled) direct effect of the treatment on the outcome, and $\text{AIE}(x, x^*) =$ $\alpha\beta(x-x^*)$ the average indirect effect. Under assumptions, these causal estimands can be identified from the observed data (Imai *and others*, 2010; VanderWeele, 2015).

With multiple mediators, a challenge is to delineate the causal structure among the mediators. When the mediators maintain the ordering information, one can directly penalize each causal connection in the directed acyclic graph (Shojaie and Michailidis, 2010). However, considering brain activity as the mediators, functional magnetic resonance imaging (fMRI) is not sufficiently informative to determine the causal ordering of the brain regions, given its low temporal resolution and high noise. To address this issue Huang and Pan (2016) introduced the concept of the principal components of the mediators, which can be used to linearly combine the candidate mediators. Assuming orthogonality, the mediation principal components are conditionally independent given the treatment. The complex causal structure in Figure 1b can be transformed into a problem with parallel causal mediation pathways, as shown in Figure 1c. As discussed in Imai and Yamamoto (2013) and VanderWeele (2015), with causally independent multiple mediators, it is equivalent to performing a series of marginal mediation analyses.

Let $X$ denote the treatment assignment, $\mathbf{M}(x) = (M_1(x), \ldots, M_p(x))^\top \in \mathbb{R}^p$ the $p$-dimensional potential outcome of mediator given the treatment assignment at level $x$. Let $\tilde{M}^{(j)}(x) = \mathbf{M}(x)^\top \boldsymbol{\phi}_j$ ($j = 1, \ldots, p$) be a linear projection of the potential outcome $\mathbf{M}(x)$, such that

$$\tilde{M}^{(j)}(x) \perp\!\!\!\perp \tilde{M}^{(k)}(x) \mid X = x, \quad \text{for } j \neq k; \tag{2.4}$$

that is, the mediators in the projection space are causally independent under the definition in Imai and Yamamoto (2013). In this setting, the problem is equivalent to conducting a series of marginal mediation analyses. For subject $i$ ($i = 1, \ldots, n$), under the LSEM framework, for each $j = 1, \ldots, p$,

$$\tilde{M}_i^{(j)} = \alpha_{0j} + \alpha_j X_i + \xi_{ij}, \tag{2.5}$$

$$Y_i = \beta_{0j} + \gamma_j X_i + \beta_j \tilde{M}_i^{(j)} + \eta_{ij}, \tag{2.6}$$

where $\{\alpha_{0j}, \alpha_j, \beta_{0j}, \gamma_j, \beta_j\}$ is the model parameter, and $\xi_{ij}$ and $\eta_{ij}$ are independent model errors normally distributed with mean zero. Here $\alpha_j \beta_j (x - x^*)$ is the average indirect effect of the projected mediator $\tilde{M}^{(j)}$ comparing treatment $x$ and $x^*$, and $\sum_{j=1}^{p} \alpha_j \beta_j (x - x^*)$ is the total average indirect effect. Under this marginal LSEM, $\gamma_j (x - x^*)$ is interpreted as the treatment effect not mediated through the mediator $\tilde{M}^{(j)}$.

As proposed in Huang and Pan (2016), obtaining these causally independent mediators is achieved through principal component analysis. Consider model

$$M_{ij} = \tau_{0j} + \tau_j X_i + \epsilon_{ij}, \quad \text{for } i = 1, \ldots, n, \quad j = 1, \ldots, p, \tag{2.7}$$

where $\{\tau_{0j}, \tau_j\}$ are model coefficients, and $\epsilon_{ij}$ is the normally distributed random error with mean zero. Let $\boldsymbol{\epsilon}_i = (\epsilon_{i1}, \ldots, \epsilon_{ip})^\top$, assume

$$\mathbb{E}\left(\boldsymbol{\epsilon}_i \boldsymbol{\epsilon}_i^\top\right) = \boldsymbol{\Sigma} = \boldsymbol{\Phi} \boldsymbol{\Lambda} \boldsymbol{\Phi}^\top, \tag{2.8}$$

where $\boldsymbol{\Phi} = (\boldsymbol{\phi}_1, \ldots, \boldsymbol{\phi}_p) \in \mathbb{R}^{p \times p}$ is an orthonormal matrix such that $\boldsymbol{\Phi}^\top \boldsymbol{\Phi} = \mathbf{I}$, and $\boldsymbol{\Lambda} = \text{diag}\{\lambda_1, \ldots, \lambda_p\}$ is a $p$-dimensional diagonal matrix, and thus $\boldsymbol{\Phi} \boldsymbol{\Lambda} \boldsymbol{\Phi}^\top$ is the spectral decomposition of the positive-definite matrix $\boldsymbol{\Sigma}$. The columns of $\tilde{\mathbf{M}} = \mathbf{M} \boldsymbol{\Phi}$ are conditionally independent given $X$, where $\mathbf{M} = (\mathbf{M}_1, \ldots, \mathbf{M}_n)^\top$ and $\mathbf{M}_i = (M_{i1}, \ldots, M_{ip})^\top$ for $i = 1, \ldots, n$. The method proposed in Huang and Pan (2016) is summarized as follows:

**Step 1.** For $j = 1, \ldots, p$, fit model (2.7) and denote the residuals as $\{e_{i1}, \ldots, e_{ip}\}$ for $i = 1, \ldots, n$.

**Step 2.** Conduct PCA on the residuals $\{e_{i1}, \ldots, e_{ip}\}_{i=1}^n$ to obtain $\hat{\boldsymbol{\Phi}} = (\hat{\boldsymbol{\phi}}_1, \ldots, \hat{\boldsymbol{\phi}}_p)$ and $\hat{\boldsymbol{\Lambda}}$.

**Step 3.** Let $\tilde{M}_i^{(j)} = \mathbf{M}_i \hat{\boldsymbol{\phi}}_j$. Using the transformed mediators, perform marginal mediation analysis using models (2.5) and (2.6), for $j = 1, \ldots, q$, where analogous to PCA, $q$ is determined by the designated proportion of variance explained.

Though the focus of Huang and Pan (2016) is on hypothesis testing for the direct and total indirect effects, one can adapt their approach to make inference about individual pathway effects.

It is well-known that PC loadings are sign nonidentifiable. Here, we show that though the estimate of $\alpha$ and $\beta$ are not sign identifiable, the estimate of the indirect and direct effects are sign consistent.

PROPOSITION 1  Let $\tilde{\mathbf{M}}^{(1j)} = \mathbf{M}\boldsymbol{\phi}_j$, $\tilde{\mathbf{M}}^{(2j)} = \mathbf{M}(-\boldsymbol{\phi}_j) = -\mathbf{M}\boldsymbol{\phi}_j$. Let $(\hat{\alpha}_j^{(s)}, \hat{\beta}_j^{(s)}, \hat{\gamma}_j^{(s)})$ denote the estimate from the transformed mediator $\tilde{\mathbf{M}}^{(sj)}$ using models (2.5) and (2.6), for $s = 1, 2$. Then

$$\hat{\alpha}_j^{(1)} = -\hat{\alpha}_j^{(2)}, \quad \hat{\beta}_j^{(1)} = -\hat{\beta}_j^{(2)}, \quad \hat{\gamma}_j^{(1)} = \hat{\gamma}_j^{(2)}.$$

Thus the estimated direct and indirect (estimated by the product $\alpha\beta$) effects are sign invariant.

## 3. SPARSE HIGH-DIMENSIONAL MEDIATION ANALYSIS

### 3.1 *Motivation*

As discussed in Huang and Pan (2016), the estimated causal effects "do not necessarily have an intuitive interpretation", since the transformed mediators are linear combinations of the original mediators. This drawback commonly occurs in PCA-based studies. An informal way to reduce the number of variables is to set a hard threshold and force the loadings with absolute value below the threshold to be zero. However, this can be potentially misleading; for example, see Cadima and Jolliffe (1995). Jolliffe *and others* (2003) introduced the modified PCA based on the Lasso (Tibshirani, 1996) to yield possible zero loadings. This sparse PCA framework was then further studied by Zou *and others* (2006) based on the fact that sparsifying the PC loadings is equivalent to a regression-type optimization problem. In this study, we propose a sparse PCA based mediation analysis approach to estimate the mediator PCs with sparse loadings.

### 3.2 *The Lasso, the generalized Lasso and the elastic net*

The least absolute shrinkage and selection operator (Lasso) was introduced by Tibshirani (1996) to perform simultaneous variable selection and estimation in linear regression. Let $Y = (Y_1, \ldots, Y_n)^\top$ denote the dependent variable, and $\mathbf{X} = (\mathbf{X}_1, \ldots, \mathbf{X}_n)^\top$ where $\mathbf{X}_i = (X_{i1}, \ldots, X_{ip})^\top$ ($i = 1, \ldots, n$) the design matrix with $p$ predictors. The lasso solution minimizes the squared-error loss under $\ell_1$ regularization. That is,

$$\hat{\beta}_{\text{lasso}} = \arg\min_{\beta \in \mathbb{R}^p} \|Y - \mathbf{X}\beta\|_2^2 + \lambda\|\beta\|_1, \tag{3.9}$$

where $\beta \in \mathbb{R}^p$ is the model coefficient, $\lambda \geqslant 0$ is a tuning parameter, and $\|\mathbf{x}\|_1 = \sum_{j=1}^p |x_j|$ is the $\ell_1$-norm of a $p$-dimensional vector $\mathbf{x} \in \mathbb{R}^p$. When the tuning parameter $\lambda$ is large enough, some coefficients will be shrunk to exactly zero. Under certain regularity conditions, the Lasso estimator has been shown to be both consistent and sparsistent (see Meinshausen and Bühlmann, 2006; Wainwright, 2009; Zhao and Yu, 2006).

Tibshirani and Taylor (2011) considered the problem of the generalized lasso to enforce structured constraints instead of pure sparsity. The problem can be formalized as

$$\hat{\beta}_{\text{glasso}} = \arg\min_{\beta \in \mathbb{R}^p} \|Y - \mathbf{X}\beta\|_2^2 + \lambda\|\mathbf{D}\beta\|_1, \tag{3.10}$$

where $\mathbf{D} \in \mathbb{R}^{m \times p}$ is a prespecified penalty matrix. The asymptotic properties of the solution was studied in She (2010).

The lasso has several limitations. As discussed in Zou and Hastie (2005), one limitation is that for predictors with high collinearity, the lasso tends to randomly select one of them; and second, when $p > n$, the lasso selects at most $n$ variables. In the mediation analysis setting, the mediators are potentially causally dependent, which violates the incoherence assumption for the lasso. Considering brain voxels as mediators, where $p \sim 100,000$, with limited number of trials $n < 100$, it is not desirable to select at most $n$ voxels. Zou and Hastie (2005) introduced the elastic net to address these drawbacks by introducing a convex combination of $\ell_1$ and $\ell_2$ penalties. The elastic net solution is written as

$$\hat{\beta}_{\text{en}} = (1 + \lambda_2)\left\{\arg\min_{\beta \in \mathbb{R}^p} \|Y - \mathbf{X}\beta\|_2^2 + \lambda_1\|\beta\|_1 + \lambda_2\|\beta\|_2^2\right\}, \tag{3.11}$$

where $\lambda_1, \lambda_2 \geqslant 0$. When $\lambda_2$ is positive, the elastic net approach can potentially choose all the variables and overcomes the drawbacks with the $\ell_1$ penalty only. In this study, we consider the following generalized elastic net solution, i.e.,

$$\hat{\beta}_{\text{gen}} = (1 + \lambda_2) \left\{ \underset{\beta \in \mathbb{R}^p}{\arg\min} \|Y - \mathbf{X}\beta\|_2^2 + \lambda_1 \|\mathbf{D}\beta\|_1 + \lambda_2 \|\beta\|_2^2 \right\}, \tag{3.12}$$

to impose a structured regularization.

### 3.3 *Sparse approximation*

For PCA, Zou *and others* (2006) studied a simple regression approach to recover the PC loadings and showed that with ridge penalty, the normalized solution to the regression problem by regressing the loadings on the variables is independent of $\lambda$. With this property, the inclusion of ridge penalty is not meant to penalize the regression coefficients but to ensure the reconstruction of principal components. As described in Section 2, in mediation analysis, PCA is conducted on the residuals of the mediation models. Therefore, we propose to sparsify the loadings using these model residuals, i.e., considering the following optimization problem, for $k = 1, \ldots, q$,

$$\hat{\mathbf{v}}_k = (1 + \lambda_2) \left\{ \underset{\mathbf{v} \in \mathbb{R}^p}{\arg\min} \|\tilde{\mathbf{e}}^{(k)} - \mathbf{E}\mathbf{v}\|_2^2 + \lambda_1 \|\mathbf{D}\mathbf{v}\|_1 + \lambda_2 \|\mathbf{v}\|_2^2 \right\}, \tag{3.13}$$

where $\mathbf{v} = (v_1, \ldots, v_p)^\top \in \mathbb{R}^p$; $\mathbf{E} = (\mathbf{e}_1, \ldots, \mathbf{e}_n)^\top$ with $\mathbf{e}_i = (e_{i1}, \ldots, e_{ip})^\top$ is the residual matrix, and $\tilde{\mathbf{e}}^{(k)} = \mathbf{E}\hat{\boldsymbol{\phi}}_k$ is the $k$th principal component. Then $\hat{\mathbf{w}}_k = \hat{\mathbf{v}}_k / \|\hat{\mathbf{v}}_k\|_2$ is a sparse approximation of $\hat{\boldsymbol{\phi}}_k$.

In neuroimaging studies, a spatial smoothness constraint is commonly applied (for example, see Grosenick *and others* (2013); Liu *and others* (2018)). In this study, we consider a fused lasso penalty (Tibshirani *and others*, 2005) to impose a local constancy of the PC profile:

$$\hat{\mathbf{v}}_k = (1 + \lambda_2) \left\{ \underset{\mathbf{v} \in \mathbb{R}^p}{\arg\min} \|\tilde{\mathbf{e}}^{(k)} - \mathbf{E}\mathbf{v}\|_2^2 + \lambda_{11} \|\mathbf{v}\|_1 + \lambda_{12} \sum_{(j,j') \in \mathcal{E}} |v_j - v_{j'}| + \lambda_2 \|\mathbf{v}\|_2^2 \right\}, \tag{3.14}$$

where $\mathcal{E}$ is an edge set such that $(j, j') \in \mathcal{E}$ if $M_j$ and $M_{j'}$ are "neighbor" brain voxels/regions.

Formulation (3.14) is a special case of the generalized elastic net (3.12) with $\mathbf{D}$ the fused lasso matrix corresponding to the underlying graph with edge set $\mathcal{E}$.

3.3.1 *Tuning parameter selection* Zou *and others* (2006) showed that the inclusion of a ridge penalty does not penalize the regression coefficients. Thus, when $n \gg p$, we can set the tuning parameter for ridge penalty $\lambda_2$ to zero; and when $n \ll p$, we can in principle use any positive $\lambda_2$. The objective of sparse PCA is to sparsify the loadings while preserving the proportion of variance explained. Zou *and others* (2006) proposed to choose $\lambda_1$ by examining the trace plot of the percentage of total variance explained calculated from the adjusted total variance. We apply the same strategy to choose tuning parameters $\lambda_{11}$ and $\lambda_{12}$ in (3.14).

### 3.4 *Mediation analysis with sparse principal components*

In this section, we discuss the analysis with the sparse PC of the mediators. Let $\check{M}_i^{(k)} = \mathbf{M}_i^\top \mathbf{w}_k = \sum_{j=1}^p M_{ij} w_{kj}$, and for $k = 1, \ldots, q$, fit models

$$\check{M}_i^{(k)} = a_0 + a_k X_i + \check{\epsilon}_i^{(k)},$$
$$Y_i = b_0 + cX_i + b_k \check{M}_i^{(k)} + \check{\eta}_i^{(k)}, \tag{3.15}$$

where $\check{\epsilon}_i^{(k)}$ and $\check{\eta}_i^{(k)}$ are random errors normally distributed with mean zero. One advantage of the PCA mediation analysis is that the transformed mediator PCs are conditionally independent, and fitting the LSEM with multiple mediators is equivalent to using marginal LSEMs for each individual mediator. By sparsifying the loading vector, the orthogonal constraint is not explicitly imposed. To achieve the conditional independence, we include a regression projection step to remove the conditional linear dependence between the transformed mediators which is analogous to the procedure proposed in Zou *and others* (2006).

Let $\check{M}_i^{(k \cdot 1, \ldots, k-1)}$ denote the residual after adjusting for $\check{M}_i^{(1)}, \ldots, \check{M}_i^{(k-1)}$ when controlling

$X_i$ (for $i = 1, \ldots, n$), that is

$$\check{M}_i^{(k \cdot 1, \ldots, k-1)} = \check{M}_i^{(k)} - \check{\mathbf{M}}_i^{(1, \ldots, k-1)\top} \hat{\mathbf{\Pi}}_{1, \ldots, k-1} \quad (3.16)$$

where $\check{\mathbf{M}}_i^{(1, \ldots, k-1)} = (\check{M}_i^{(1)}, \ldots, \check{M}_i^{(k-1)})^\top$, and $\hat{\mathbf{\Pi}}_{1, \ldots, k-1}$ is the estimated coefficient in model

$$\check{M}_i^{(k)} = \pi_0 + \pi_1 X_i + \check{\mathbf{M}}_i^{(1, \ldots, k-1)\top} \mathbf{\Pi}_{1, \ldots, k-1} + \tau_i^{(k)}, \quad (3.17)$$

where $\tau_i^{(k)}$ is normally distributed model error with mean zero. The new mediators $\check{M}^{(k \cdot 1, \ldots, k-1)}$ are uncorrelated given the treatment $X$, thus we can use model (3.15) to estimate the indirect effect of each individual mediation pathway.

We summarize the steps of mediation analysis with sparse principal components in Algorithm 1. To perform inference of model parameters, we propose a bootstrap procedure.

(i) Generate a bootstrap sample $(X_i^*, \check{M}_i^{(k \cdot 1, \ldots, k-1)*}, Y_i^*)$ of size $n$ by resampling the data with replacement, where $\check{M}_i^{(k \cdot 1, \ldots, k-1)}$ is the modified mediator PC obtained in Step 4 of Algorithm 1.

(ii) Estimate the model parameters in (3.15) using the bootstrap sample.

(iii) Repeat steps (i)-(ii) $B$ times.

Bootstrap confidence intervals can be then calculated by either percentile or bias-corrected approach (Efron, 1987) under prespecified significance level.

## 4. SIMULATION STUDY

Simulation study is conducted to examine the performance of the proposed sparse PC based mediation analysis approach. Description of the study and the results are presented in Section C of the supplementary material available at *Biostatistics* online.

---

**Algorithm 1** Mediation analysis with sparse principal components.

**Step 1.** For $j = 1, \ldots, p$, fit model (2.7) and denote the residuals as $\{e_{i1}, \ldots, e_{ip}\}$ for $i = 1, \ldots, n$.

**Step 2.** Conduct PCA on the residuals $\{e_{i1}, \ldots, e_{ip}\}_{i=1}^{n}$ and estimate the loading matrix as $\hat{\mathbf{\Phi}} = (\hat{\boldsymbol{\phi}}_1, \ldots, \hat{\boldsymbol{\phi}}_p)$.

**Step 3.** For $k = 1, \ldots, q$, let $\tilde{\mathbf{e}}^{(k)} = \mathbf{E}\hat{\boldsymbol{\phi}}_k$, where $\mathbf{E} = (\mathbf{e}_1, \ldots, \mathbf{e}_n)^\top$ and $\mathbf{e}_i = (e_{i1}, \ldots, e_{ip})^\top$ ($i = 1, \ldots, n$). Perform regularized regression using the generalized elastic net penalty (3.12) and attain the estimator $\hat{\mathbf{v}}_k(\hat{\lambda}_k)$ and its normalization $\hat{\mathbf{w}}_k(\hat{\lambda}_k) = \hat{\mathbf{v}}_k(\hat{\lambda}_k)/\|\hat{\mathbf{v}}_k(\hat{\lambda}_k)\|_2$, for $k = 1, \ldots, q$, where $\hat{\lambda}_k$ is chosen based on the method discussed in Section 3.3.1.

**Step 4.** Let $\check{M}_i^{(k)} = \mathbf{M}_i^\top \hat{\mathbf{w}}_k$ for $i = 1, \ldots, n$. For $k = 2, \ldots, q$ obtain the modified $\check{M}_i^{(k \cdot 1, \ldots, k-1)}$.

**Step 5.** Fit model (3.15) using the causally independent $\{\check{M}_i^{(1)}, \check{M}_i^{(2 \cdot 1)}, \ldots, \check{M}_i^{(q \cdot 1, \ldots, q-1)}\}$ to yield estimates of the causal effects.

---

## 5. A Task-Based Functional MRI Study

We analyze a task-based fMRI study using the proposed sparse principal component of mediation approach. The data set is downloaded from the OpenfMRI database (accession number ds000002). In the experiment, participants were instructed to perform a probabilistic classification learning (PCL) task using "weather prediction" (Aron *and others*, 2006). To avoid inter-subject heterogeneity, we use the data from a single healthy right-handed English-speaking subject aged between 21 to 26. The experiment consisted of $n = 80$ trails with ten cycles, and within each there are five PCL trails intermixed with three baseline trails. Under the weather prediction trial, a visual stimulus was presented at a randomized location. The participant would respond by pressing either the left button for a "sun" prediction or the right button for a "rain" prediction. The experiment included baseline trials to control for visual stimulation, button press and computer response to button press. In this study, reaction time ($Y$) is considered as the outcome of interest.

The goal is to discover the brain networks that have an intermediate effect on the reaction time when comparing PCL ($X = 1$) and baseline ($X = 0$) trials.

180 functional T2$^*$-weighted echoplanar images (EPI) (4 mm slice thickness, 33 slices, TR = 2 s, TE = 30 ms, flip angle = 90°, matrix 64 × 64, field of view 200) were acquired from a 3 T Siemens Allegra MRI scanner. For registration purposes, a matched-bandwidth High-Resolution scan (same slice prescription as EPI) and MPRAGE (TR = 2.3, TE = 2.1, FOV = 256, matrix = 192 × 192, saggital plane, slice thickness = 1 mm, 160 slices) were acquired for each participant. Preprocessing for both anatomical and functional images was conducted using Statistical Parametric Mapping version 5 (SPM5) (Wellcome Department of Imaging Neuroscience, University College London, London, UK), including slice timing correction, realignment, coregistration, normalization, and smoothing. The Blood-oxygen-level dependent (BOLD) time courses are extracted from $p = 264$ putative brain functional regions defined in Power *and others* (2011). These brain regions are grouped into eleven (ten functional and one uncertain) modules. We first employ the general linear model approach to acquire a summary measure of brain activity for each trial at each region (Atlas *and others*, 2010; Lindquist, 2008; Rissman *and others*, 2004). These single-trial brain activities are used as the mediators ($M$).

We apply the PCA based high-dimensional mediation analysis proposed in Huang and Pan (2016) and our proposed sparse PCA based method. When conducting sparse approximation, we consider a fused lasso penalty (3.14), where edge set $\mathcal{E}$ is defined based on the spatial location of the brain region as well as the module information. If brain region $j$ and $k$ are from the same functional module, then $(j, k) \in \mathcal{E}$; if region $k$ is the nearest neighbor of region $j$, then $(j, k) \in \mathcal{E}$. In the PCA based analysis, the first 18 PCs, which account for 76.2% of the total variation, are tested for mediation effect (see Figure B.1 in the supplementary material). PC3 shows significant positive indirect effect, where the linear combination of all brain regions shows deactivation in PCL compared to baseline ($\alpha$ estimate $-0.376$ with 95% confidence interval $(-0.592, -0.160)$), and this deactivation further increases the reaction time ($\beta$ estimate $-0.235$ with 95% confidence interval $(-0.365, -0.105)$). Figure 2 shows the sparse approximation under the fused lasso penalty. The estimated model coefficients, as well as the indirect effect are very close to those in the PCA based analysis (Table 1 and Figure B.1 in the supplementary material). From the figure, the whole cerebellum module is regularized to zero. All the visual modules, including lateral, medial and occipital pole visual, yield negative loadings; the the positive loadings are mainly from the auditory, default mode network, executive control and frontoparietal modules (see Figure B.3 in the supplementary material). Figure 3 shows the regions with positive and negative loadings in a brain map. The medial frontal and parietal cortex and the auditory cortex are often identified to be deactivated (positive loadings with negative $\alpha$ estimate) when the task involves visual stimuli (Poldrack *and others*, 2001; Aron *and others*, 2004). The positive loading map (Figure 3a) also includes the medial temporal lobe (MTL) regions, which is in line with the findings in the existing literature (Poldrack *and others*, 2001). The negative loading map (Figure 3b) consists of the visual cortex and the basal ganglia (caudate nucleus). This classification learning task is a nondeclarative memory procedure. The opposite sign of the loadings in MTL and striatum verifies the competing role of these two memory system regions during learning (Poldrack *and others*, 2001).

## 6. Discussion

In this study, we introduce a sparse principal component analysis (PCA) based approach to perform mediation analysis with high-dimensional mediators. As an extension of the method introduced in Huang and Pan (2016), the proposed approach enables meaningful interpretations about the transformed mediators by employing the sparse PCA method proposed in Zou *and others* (2006). In the task-based fMRI application, we consider the fused lasso penalty based on spatial information to enforce local smoothness and constancy. With sparse loadings, the activation patterns of the brain regions in the PC with significant mediation effect are consistent with those found in the existing literature.

Though our proposed high-dimensional mediation analysis approach is motivated by neuroimaging studies, the fundamental principle can be generalized to other areas, for example genetics studies. Given the spatial information of the brain, we consider a special type of the generalized lasso, that is the fused lasso. Other lasso-type structured regularization can be adopted based on data characteristics, including group lasso (Yuan and Lin, 2006) and its variations (Yuan and others, 2011).

## 7. Software

Software in the form of R code, together with a sample input data set and complete documentation is available on Github at `https://github.com/zhaoyi1026/spcma`.

## 8. Supplementary Material

Supplementary material is available online at `http://biostatistics.oxfordjournals.org`.

## Acknowledgments

*Conflict of Interest*: None declared.

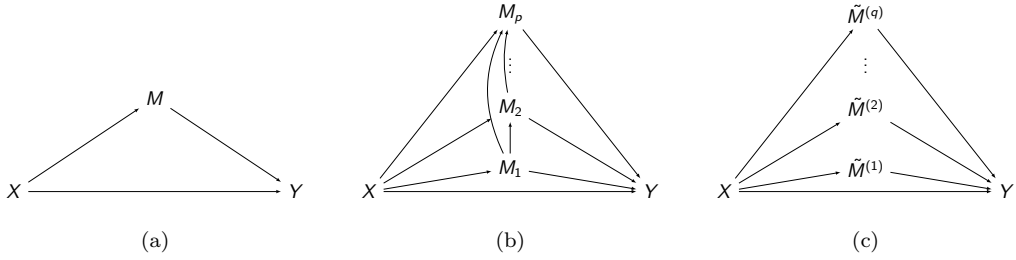

Fig. 1. Causal diagram of (a) single mediator, (b) $p$ dependent ordered mediators and (c) $q$ causally independent transformed mediators.

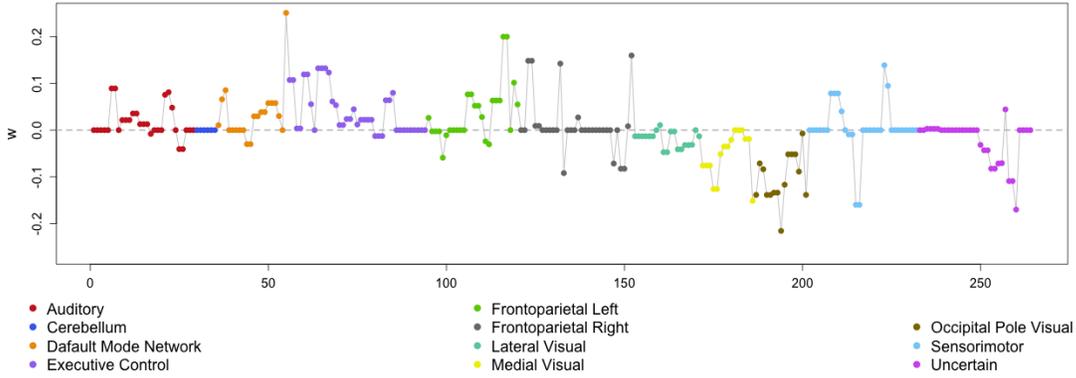

Fig. 2. The sparse approximation of the loadings of PC3.

Table 1. *The estimate (Est.) of model parameters $\alpha$, $\beta$, and the indirect effect (IE), as well as the 95% bootstrap confidence interval (CI) of PC3, which yields significant IE.*

|  |  | $\alpha$ |  | $\beta$ |  | IE ($\alpha\beta$) | |
|---|---|---|---|---|---|---|---|
|  |  | Est. | 95% CI | Est. | 95% CI | Est. | 95% CI |
| PC3 | PCA | -0.376 | (-0.592, -0.160) | -0.235 | (-0.365, -0.105) | 0.089 | (0.020, 0.176) |
|  | SPCA | -0.343 | (-0.534, -0.151) | -0.262 | (-0.410, -0.115) | 0.090 | (0.025, 0.178) |

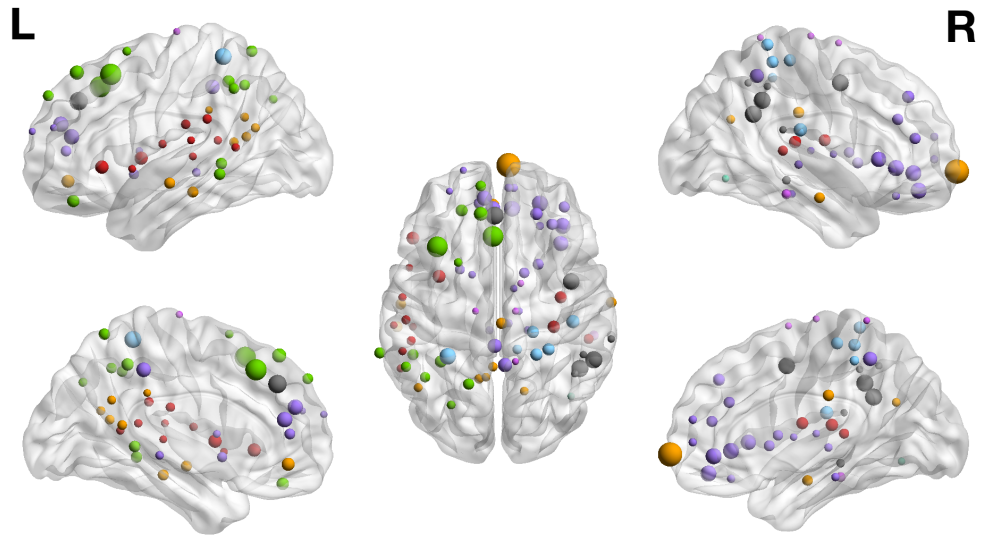

(a) Brain regions with positive loadings.

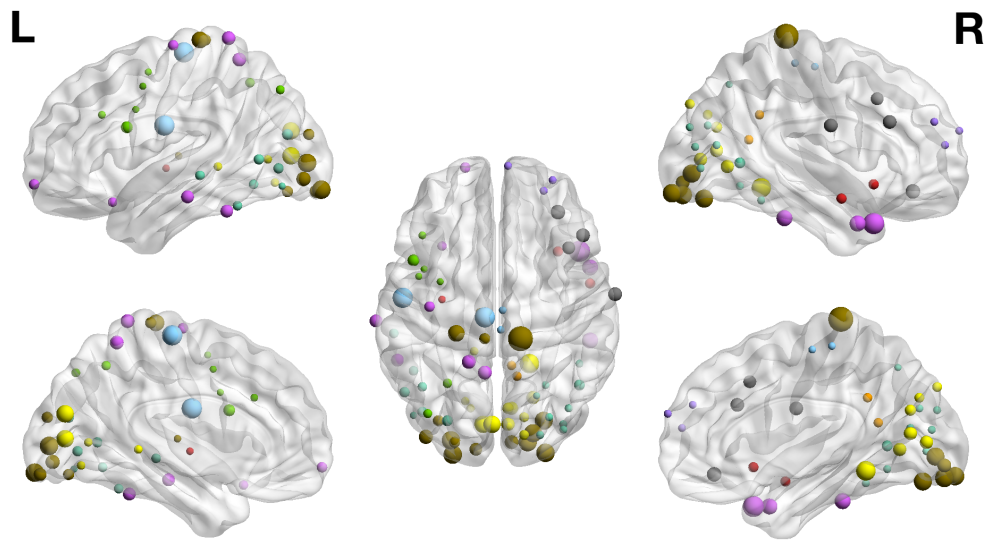

(b) Brain regions with negative loadings

Fig. 3. Brain regions with (a) positive and (b) negative loadings in sparsified PC3. The module color code is the same as in Figure 2. The size of the node is proportion to the absolute value of the loading.

# Sparse Principal Component based High-Dimensional Mediation Analysis Supplementary Materials

## APPENDIX

## A. Theory and Proof

### A.1 *Proof of Proposition 1*

*Proof.* Without loss of generality, we assume the data is center at zero and drop the intercept term in models (2.5) and (2.6). The ordinary least square (OLS) estimator of the model coefficient in model (2.5) is

$$\hat{\alpha}_j^{(1)} = (\mathbf{X}^\top \mathbf{X})^{-1} \mathbf{X}^\top \tilde{\mathbf{M}}^{(1j)} = (\mathbf{X}^\top \mathbf{X})^{-1} \mathbf{X}^\top \mathbf{M} \boldsymbol{\phi}_j,$$

$$\hat{\alpha}_j^{(2)} = (\mathbf{X}^\top \mathbf{X})^{-1} \mathbf{X}^\top \tilde{\mathbf{M}}^{(2j)} = (\mathbf{X}^\top \mathbf{X})^{-1} \mathbf{X}^\top \mathbf{M}(-\boldsymbol{\phi}_j) = -(\mathbf{X}^\top \mathbf{X})^{-1} \mathbf{X}^\top \mathbf{M} \boldsymbol{\phi}_j,$$

$$\Rightarrow \quad \hat{\alpha}_j^{(1)} = -\hat{\alpha}_j^{(2)}.$$

Assume $\tau_1^2 = \mathbf{X}^\top \mathbf{X}$, $\tau_2^2 = \tilde{\mathbf{M}}^{(1j)\top}\tilde{\mathbf{M}}^{(1j)} = \boldsymbol{\phi}^\top \mathbf{M}^\top \mathbf{M} \boldsymbol{\phi}_j = \tilde{\mathbf{M}}^{(2j)\top}\tilde{\mathbf{M}}^{(2j)}$, $\rho = \mathbf{X}^\top \tilde{\mathbf{M}}^{(1j)} = \mathbf{X}^\top \mathbf{M} \boldsymbol{\phi}_j$, then $\mathbf{X}^\top \tilde{\mathbf{M}}^{(2j)} = -\rho$. The OLS estimator of $\beta_j$ and $\gamma_j$ in model (2.6) are

$$\begin{pmatrix}\hat{\gamma}_j^{(1)}\\ \hat{\beta}_j^{(1)}\end{pmatrix} = \frac{1}{\tau_1^2\tau_2^2 - \rho^2}\begin{pmatrix}\tau_2^2\mathbf{X}^\top Y - \rho\tilde{\mathbf{M}}^{(1j)\top}Y\\ \tau_1^2\tilde{\mathbf{M}}^{(1j)\top}Y - \rho\mathbf{X}^\top Y\end{pmatrix} = \frac{1}{\tau_1^2\tau_2^2 - \rho^2}\begin{pmatrix}\tau_2^2\mathbf{X}^\top Y - \mathbf{X}^\top \mathbf{M}\boldsymbol{\phi}_j\boldsymbol{\phi}_j^\top \mathbf{M}^\top Y\\ \tau_1^2\boldsymbol{\phi}_j^\top \mathbf{M}^\top Y - \boldsymbol{\phi}_j^\top \mathbf{M}^\top \mathbf{X}\mathbf{X}^\top Y\end{pmatrix}$$

and

$$\begin{pmatrix}\hat{\gamma}_j^{(2)}\\ \hat{\beta}_j^{(2)}\end{pmatrix} = \frac{1}{\tau_1^2\tau_2^2 - \rho^2}\begin{pmatrix}\tau_2^2\mathbf{X}^\top Y + \rho\tilde{\mathbf{M}}^{(2j)\top}Y\\ \tau_1^2\tilde{\mathbf{M}}^{(2j)\top}Y + \rho\mathbf{X}^\top Y\end{pmatrix} = \frac{1}{\tau_1^2\tau_2^2 - \rho^2}\begin{pmatrix}\tau_2^2\mathbf{X}^\top Y - \mathbf{X}^\top \mathbf{M}\boldsymbol{\phi}_j\boldsymbol{\phi}_j^\top \mathbf{M}^\top Y\\ -\tau_1^2\boldsymbol{\phi}_j^\top \mathbf{M}^\top Y + \boldsymbol{\phi}_j^\top \mathbf{M}^\top \mathbf{X}\mathbf{X}^\top Y\end{pmatrix}.$$

$$\Rightarrow \quad \hat{\beta}^{(1j)} = -\hat{\beta}^{(2j)}, \quad \hat{\gamma}^{(1j)} = \hat{\gamma}^{(2j)}.$$

Therefore, the estimate of the indirect effect

$$\widehat{\text{IE}}^{(1j)} = \hat{\alpha}^{(1j)}\hat{\beta}^{(1j)} = \hat{\alpha}^{(2j)}\hat{\beta}^{(2j)} = \widehat{\text{IE}}^{(2j)}.$$

The estimate of the direct and indirect effects are sign invariant with respect to the loadings. □

## B. Additional fMRI Study Results

Figure B.1 shows the estimate of model coefficient and the indirect effect (IE) of the first 18 PCs in both the PCA and sparse PCA (SPCA) based mediation analysis. From the figure, only PC3 yields significant positive mediation effect with significant negative $\alpha$ and $\beta$ estimates.

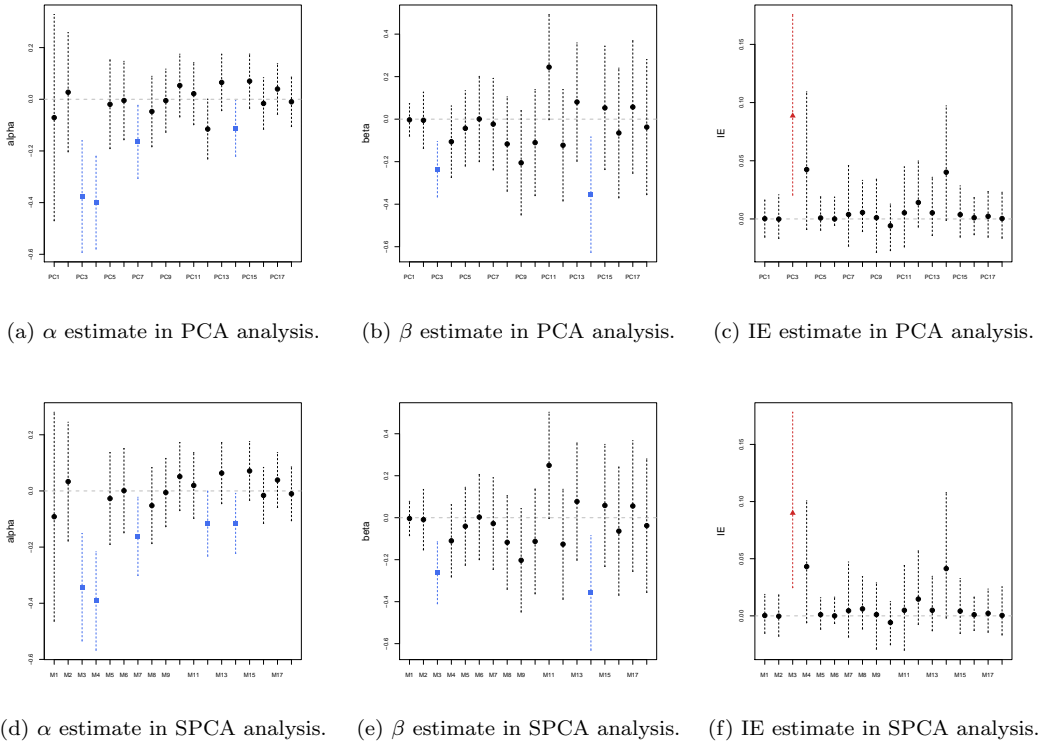

(a) $\alpha$ estimate in PCA analysis. (b) $\beta$ estimate in PCA analysis. (c) IE estimate in PCA analysis.

(d) $\alpha$ estimate in SPCA analysis. (e) $\beta$ estimate in SPCA analysis. (f) IE estimate in SPCA analysis.

Fig. B.1. Estimate of model coefficients $\alpha$ and $\beta$, and the indirect effect (IE) in the (a)&(b)&(c) PCA based and (d)&(e)&(f) sparse PCA (SPCA) based analysis of the first 18 PCs.

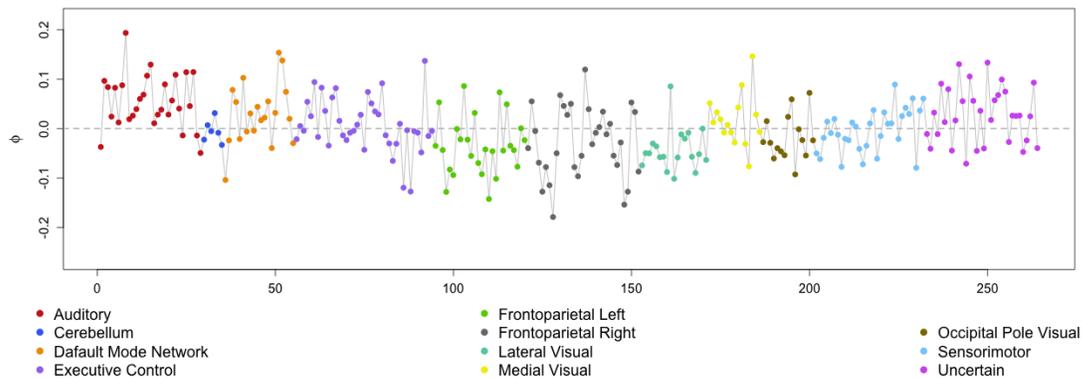

Fig. B.2. The loadings of PC3.

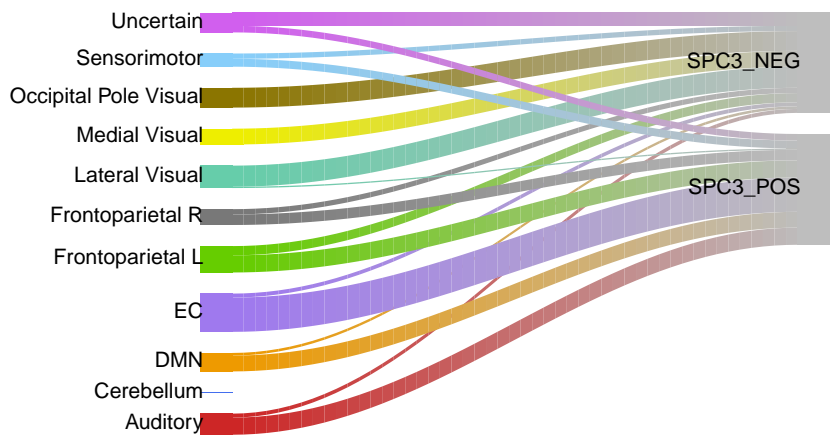

Fig. B.3. River plot of the sparse approximation of PC3. The downstream is divided into positive (POS) and negative (NEG) loadings. DMN: default mode network, EC: executive control, L: left, R: right.

Figure B.2 shows the loadings of PC3. From the figure, we do not observe any clear patterns within each functional module. Figure B.3 presents the river plot of the sparse loadings of PC3. The downstream is separated for positive and negative loadings. From the figure, the visual regions (lateral, medial and occipital pole) yield negative loadings; and the positive loadings are contributed mainly by auditory, default mode network (DMN), executive control (EC), and frontoparietal cortex. The whole cerebellum module is penalized to zero.

C. Simulation Study

The simulated data is generated following models (2.5) and (2.6), and then the mediators are linearly transformed back to the original space. In the simulation, $p = 200$. The orthogonal matrix $\boldsymbol{\Phi}$ is generated with sparse loadings, and the eigenvalues in $\boldsymbol{\Lambda}$ decay exponentially. The sample sizes are set to be $n = 50, 100, 500, 1000, 5000$ to contain both cases where $n > p$ and $n < p$. The simulation is repeated 200 times. We compare the performance of (1) the PCA based mediation analysis (PCA) and (2) the proposed sparse PC based mediation analysis (SPCA). The number of PCs is chosen so that at least 75% of the data variation is explained. The sparse PC based mediation analysis is performed following Algorithm 1.

Figure C.1 shows the estimate of model coefficients, as well as the indirect effect (IE) under different values of $n$. Since the estimates of $\alpha$ and $\beta$ are sign nonidentifiable, we compare the estimate of their absolute values. From the figure it is clear that, as the number of observations increases, the estimate from both methods converge to the truth. The PCA approach yields lower bias in estimating $|\alpha|$ and $|\beta|$, while the difference between the two methods diminishes when estimating the indirect effect.

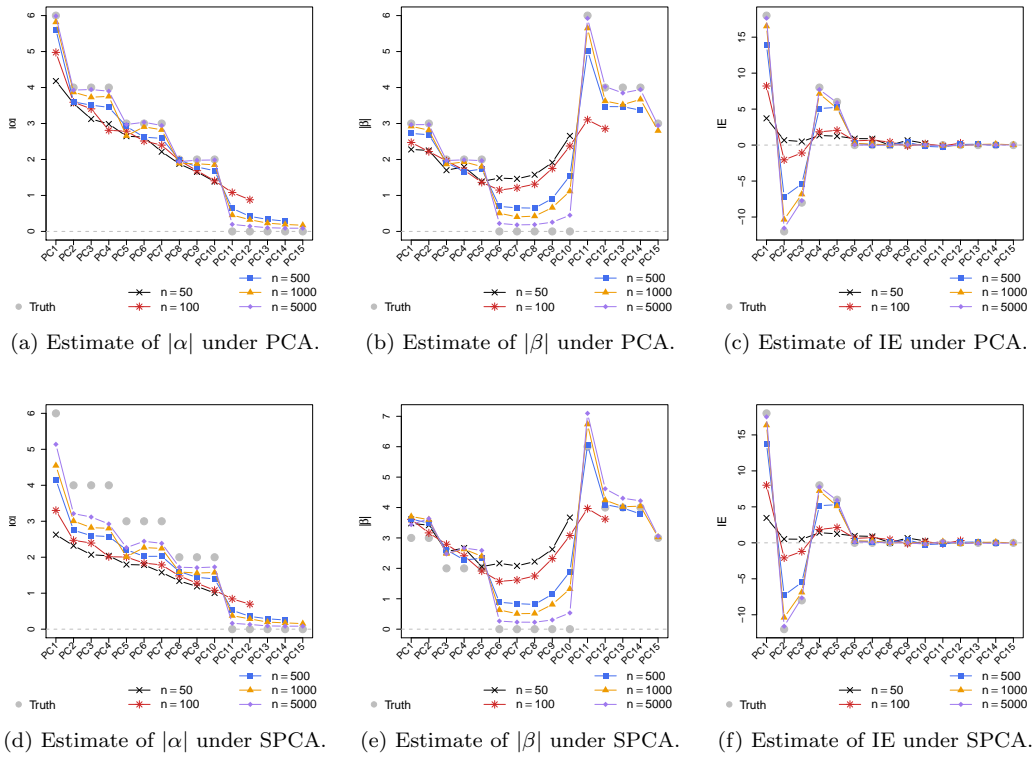

Fig. C.1. Estimate of $|\alpha|$, $|\beta|$ and the indirect effect (IE) over 200 replications under different numbers of observations.